\documentclass[11pt, nofootinbib, notitlepage, showpacs, floatfix, tightenlines]{revtex4-1} 
\usepackage{amssymb, amsmath, amsfonts, mathrsfs}
\usepackage{footnote}
\usepackage{multirow, subfigure}
\usepackage{graphicx}
\usepackage{natbib}

\def\thesection{\arabic{section}}

\begin{document}
\title{$R$-Symmetry Breaking in Supersymmetric Hybrid Inflation}

\author{Matthew Civiletti}\email{mcivil@udel.edu}
\affiliation{Bartol Research Institute, Department of Physics and Astronomy, 
University of Delaware, Newark, Delaware 19716, USA}
\author{Mansoor Ur Rehman}\email{mansoor-ur.rehman@unibas.ch}
\altaffiliation[Current address: ]{Quaid-i-Azam University, Islamabad 45320, Pakistan}
\affiliation{Bartol Research Institute, Department of Physics and Astronomy, 
University of Delaware, Newark, Delaware 19716, USA}
\author{Eric Sabo}\email{esabo@udel.edu}
\affiliation{Bartol Research Institute, Department of Physics and Astronomy, 
University of Delaware, Newark, Delaware 19716, USA}
\author{Qaisar Shafi}\email{shafi@bartol.udel.edu}
\affiliation{Bartol Research Institute, Department of Physics and Astronomy, 
University of Delaware, Newark, Delaware 19716, USA}
\author{Joshua Wickman}\email{jwickman@gccnj.edu}
\altaffiliation[Current address: ]{Gloucester County College, Sewell, New Jersey 08080, USA}
\affiliation{Bartol Research Institute, Department of Physics and Astronomy, 
University of Delaware, Newark, Delaware 19716, USA}

\begin{abstract}
We consider a supersymmetric hybrid inflation scenario in which the $U(1)$ $R$-symmetry is explicitly broken by Planck scale suppressed operators in the superpotential. We provide an example with minimal K\"ahler potential, with the $R$-symmetry breaking term relevant during inflation being $\alpha S^4$, where $S$ denotes the well-known gauge singlet inflaton superfield. The inflationary potential takes into account the radiative and supergravity corrections, as well as the soft supersymmetry breaking terms. For successful inflation, with the scalar spectral index in the currently preferred range, $n_s \approx 0.97 \pm 0.010$, $\vert \alpha \vert$ $\lesssim 10^{-7}$. The tensor to scalar ratio $r \lesssim 10^{-4}$, while $\vert \text{d} n_s / \text{d} \ln k \vert \sim \mathcal{O}(10^{-3})-\mathcal{O}(10^{-4}) $.
\end{abstract}
\pacs{98.80.Cq}

\maketitle

\section{Introduction}
Modern cosmology has seen rapid developments due to experiments such as COBE and WMAP. Augmenting their unprecedented successes will be Planck, which may for the first time yield direct evidence of inflation. At the same time, there are enormous strides being made in particle physics, with the LHC having made perhaps the greatest single discovery in the field in decades, while the testing of supersymmetry (SUSY) is highly anticipated. For the first time particle and cosmological models can be tested with precision, and the deep connections between these two fields motivate us to consider the effects that particle physics considerations have on inflationary models. This has led to models such as, among others, SUSY hybrid inflation. 

The standard version of SUSY hybrid inflation \cite{dvali1994large,  copeland1994false, [{For reviews and additional references, see: }] mazumdar2011particle, *lyth1999particle, *lazarides2001supersymmetric, *nakayama2011aspects} remains one of the most successful and well-motivated inflationary models. It is the most general non-trivial model one can write with a gauge singlet field $S$ and supermultiplets $\Phi$ and $\bar{\Phi}$ that respects $U(1)_R$, such that the latter two fields belong to non-trivial representations of the gauge group $G$. It has a connection to particle physics in that within it grand unified theories (GUTs) are naturally incorporated. In this model the gauge group $G$ is broken to a subgroup $H$ at the end of inflation, and the energy scale at which this occurs is related to local temperature anisotropies in the cosmic microwave background radiation \cite{dvali1994large}; this scale happens to be close to the GUT scale, indicating that $G$ may be related to a GUT. In addition, supergravity (SUGRA) corrections remain under control \cite{linde1997hybrid} because this model can yield solutions within the WMAP nine-year 2$\sigma$ bounds without trans-Planckian inflaton field values. In this model, where only (minimal) SUGRA and radiative corrections are added to the global potential, a numerical lower bound on the scalar spectral index, $n_s \approx 0.985$, develops. This is somewhat disfavored, although it is within the 2$\sigma$ bounds, $n_s=0.971\pm 0.010$ \cite{WMAPnine}. However, the addition of a linear soft-SUSY breaking term \cite{csenouguz2005reheat} reduces $n_s$ significantly \cite{rehman2010supersymmetric}; this term is also important in explaining the MSSM $\mu$-problem \cite{Dvali1997uq, kim2011mu, raby2012problems}. It has been shown that small gravity waves (tensor-to-scalar ratio $r$) are produced with minimal K\"ahler (retaining the lowest-order SUGRA term), radiative corrections, and positive soft mass squared and negative linear soft-SUSY breaking terms in the potential. Specifically, we know from \cite{rehman2010supersymmetric} that one can obtain $r \lesssim 10^{-5}$ in the red-tilted region in this case, although $r$ is seven orders of magnitude smaller than the upper limit at the central $n_s$ value. By contrast, the WMAP nine-year upper limit is $r < 0.13$. Planck will be able to detect $r$ of the order $\sim 10^{-2}$ \cite{PlanckBook}; if gravity waves exist at this order of magnitude, Planck will greatly reduce the number of viable inflationary models, or at the least, will rule out a considerable amount of parameter space in models in which large gravity waves can be produced.

One potential drawback of this model is that, since symmetry breaking occurs at the end of inflation, topological defects, such as magnetic monopoles and domain walls, can be produced in quantities sufficient to contradict experiment. This can be remedied by inflating along a shifted track such that the symmetry breaking precedes observable inflation \cite{civiletti2011red, jeannerot2000inflation, dvali1994large}. However, one can solve this problem in the standard case with an appropriate gauge group choice in which no topological defects are produced, such as  $G = SU(5) \times U(1)_{X}$, referred to as flipped $SU(5)$ \cite{de1980flavor, *georgi1981mu+, *barr1982new, *derendinger1984anti, *antoniadis1987supersymmetric, *shafi1999atmospheric, *[{For a review, see }] Nanopoulos2002qk}.

When doing inflation in the context of flipped $SU(5)$ \cite{kyae2006flipped, rehman2010minimal}, the $R$-charges are determined in part by the inflationary sector. By imposing a $U(1)_R$ symmetry (``$R$-symmetry"), as is standard practice, one ensures that proton decay occurs only via the six-dimensional operator (the $\mathbf{5}_h \overline{\mathbf{5}}_h$ term is disallowed). The importance of this is that we preclude rapid proton decay. However, one also prohibits terms such as the quartic couplings $\mathbf{10}_i \mathbf{10}_j \overline{\mathbf{10}}_H \overline{\mathbf{10}}_H$, which give rise to right-handed neutrino masses. (Note that the decay of Majorana right-handed neutrinos can explain the observed baryon asymmetry via leptogenesis \cite{fukugita1986barygenesis, *[{For non-thermal leptogenesis in inflation, see }] lazarides1991origin}.) If we reassign $R$-charges such that $\mathbf{10}_i \mathbf{10}_j \overline{\mathbf{10}}_H \overline{\mathbf{10}}_H$ is allowed, we end up prohibiting the Yukawa term which gives rise to down-type quark masses (${y_{ij}}^{(d)} \mathbf{10}_i \mathbf{10}_j \mathbf{5}_h$). This outcome is not necessarily catastrophic, since the relevant quark masses may be generated radiatively. On the other hand, one may invoke a ``double seesaw" mechanism \cite{deltaToverT} to account for the lack of large right-handed neutrino masses. A simpler solution to this problem is to allow higher-order (Planck scale suppressed) $R$-symmetry violating terms in the superpotential, while enforcing $R$-symmetry for renormalizable terms. With this motivation in mind, we wish to explore the effects of these additional terms on the inflationary dynamics.

In this letter, we consider the inflationary implications of such a scenario, within the context of standard SUSY hybrid inflation, which we briefly review in the Section 2. In Section \ref{sec_S4} we discuss our model including the leading nonrenormalizable terms, and in Section \ref{sec_Results} we detail our results. There, we compare our results with WMAP nine-year data, and discuss our model in light of the Planck mission.

\section{Review of the Standard Hybrid Inflation Model}\label{sec_BG}
The most general non-trivial renormalizable superpotential one can write involving a singlet superfield $S$ and two conjugate supermultiplets $\Phi$ and $\bar{\Phi}$ that preserves a gauge group $G$ and $U(1)_R$ $R$-symmetry is \cite{dvali1994large, copeland1994false}
\begin{equation} 
	W = \kappa S(\overline{\Phi} \Phi - M^{2}) ,
\label{Wstandard}
\end{equation}
where $M$ is the energy scale at which $G$ breaks and $\kappa$ is a dimensionless coupling which we take to be positive without loss of generality since we can absorb the phase of $\kappa$ into that of $S$. 
The global SUSY F-terms are given by
\begin{eqnarray}
	V_{F} \equiv \sum_i {\bigg \lvert \frac{\partial V_{\text{global}}}{\partial z_i} \bigg \rvert}^2.
\label{VFglobalgeneral}
\end{eqnarray}
Here, $z_{i}\in \{s, \phi , \overline{\phi }\}$, where $s$, $\phi$, and $\overline{\phi}$ are the scalar components of the superfields $S,\Phi$, and $\bar{\Phi}$, respectively. We choose to set the D-terms to zero by imposing $|\Phi|=|\overline{\Phi}|$, for convenience.

\begin{figure}[t]
	\begin{tabular}{cc}
		\includegraphics[width=.8\columnwidth]{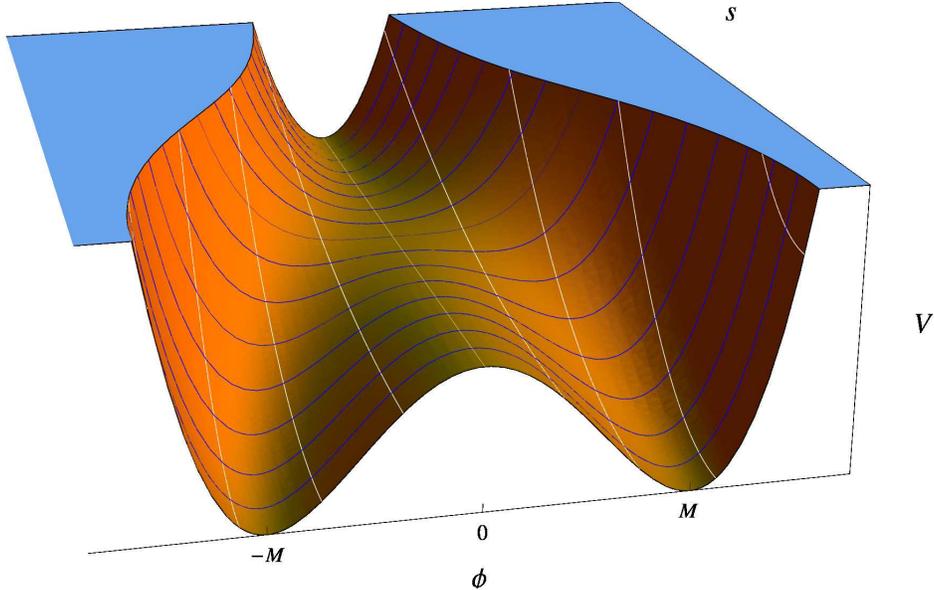}
	\end{tabular}
	\caption{The tree-level, global scalar potential $V$ in standard hybrid inflation. The variables $s$ and $\phi$ are the scalar components of the superfields $S$ and $\Phi$.}
	\label{3Dplots}
\end{figure}

Using Equation \eqref{VFglobalgeneral}, the tree-level global SUSY potential in the D-flat direction is
\begin{equation}
	V_{F}= \kappa^2\,(M^2 - \vert \phi\vert^2)^2 + 2\kappa^2 \vert s \vert^2 \vert \phi \vert^2.
\label{VFglobal}
\end{equation}
A plot of this potential in field space is shown in Figure \ref{3Dplots}. Inflation proceeds along the local minimum $\vert \phi \vert = 0$ (the inflationary track), beginning at large $\vert s \vert$ (top of Figure \ref{3Dplots}). An instability occurs at the waterfall point ${\vert \tilde{s_c} \vert}^2 = M^2$, which is the value of $\vert s \vert$ such that $0= \frac{\partial^2 V}{\partial {\vert \phi \vert}^2}\bigg \vert_{\vert \phi \vert=0}$ (the subscript ``$c$" denotes ``critical", and the symbol ${\vert \tilde{s_c} \vert}$ will be used later to denote the dimensionful inflaton field at the critical point; we maintain the same notation here for consistency.) At this point the field falls naturally into one of two global minima at ${\vert \phi \vert}^2 = M^2$. This coincides with the breaking of the gauge group $G$. At large $\vert s \vert$, the scalar potential is approximately quadratic in $\vert \phi \vert$, whereas at $\vert s \vert = 0$ Equation \eqref{VFglobal} becomes a Higgs potential.

Along the inflationary track the potential is flat ($V = \kappa^2 M^4$), and thus one cannot drive inflation. One-loop radiative corrections (RC), which should be added for consistency in any case (since SUSY is broken during inflation), can be used to drive inflation. SUSY is restored after inflation, when the field evolves to one of its global minima (where $V = 0$). The radiative corrections \cite{coleman1973radiative} involve the function
\begin{equation*}
	F(x) = \frac{1}{4} \left[ (x^4 + 1) \ln\frac{(x^4 - 1)}{x^4} + 2 x^2 \ln\frac{x^2 + 1}{x^2 - 1} + 2 \ln\left( \frac{\kappa^2 m^2 x^2}{Q^2} \right) - 3 \right], 
\label{Vloop}
\end{equation*}
where $x = |s| / |\tilde{s_c}|$ is a convenient reparametrization of the inflaton field, $\mathcal{N}$ is the dimensionality of the representation of the fields $\Phi$ and $\overline{\Phi }$, and $Q$ is the renormalization scale.  


\section{Planck Scale Suppressed $R$-Symmetry Violation}\label{sec_S4}
We now wish to determine the effects of allowing $R$-symmetry violation beyond the renormalizable level. First, let us list which additional terms one can consider in this type of model. The three lowest-order nonrenormalizable $R$-violating terms one can write with the aforementioned superfields (respecting gauge symmetry) are
\begin{eqnarray*}
	\frac{\alpha}{m_P} {S^4} \quad, \quad \frac{\beta}{m_P}{S^2}(\bar{\Phi} \Phi) \quad, \quad \frac{\gamma}{m_P}(\bar{\Phi} \Phi)^2,
\end{eqnarray*}
where $\alpha$, $\beta$, and $\gamma$ are dimensionless, and are sufficiently small such that each term is a perturbation about the standard case. Further, $m_P=M_P/\sqrt{8\pi} \approx 2.4\times 10^{18}$ GeV is the reduced Planck mass. Along the inflationary track, only the first term will lift the potential. Therefore, in this letter we consider solely the inflationary ramifications of the $S^4$ term, so that our superpotential is
\begin{equation}\label{Walphaterm}
	W = \kappa S(\overline{\Phi} \Phi - M^{2}) + \frac{\alpha}{m_P}S^4.
\end{equation}
Other inflationary tracks may be available via the inclusion the $\beta$ and $\gamma$ terms. Inclusion of these terms may lead to a form of shifted inflation; however, we do not discuss this here.

It is important to ask the following question: Can we drive inflation with this $S^4$ term, without radiative corrections nor any additional terms? Let us compute the global, tree-level scalar potential along the inflationary track. We do this via Equation \eqref{VFglobalgeneral}, which yields the dimensionless potential ($\mathcal{V} \equiv V/{m_P}^4$)
\begin{align}\label{VGlobal}
	\mathcal{V}_F\bigg|_{\vert \phi \vert = 0} = \kappa^2 m^4 - 8 \cos (\theta_\alpha + 3 \theta_s) \vert \alpha \vert \, x^3 \kappa {s_c}^3 m^2 + 16 x^6 \, {\vert \alpha \vert}^2 \, {s_c}^6,
\end{align}
where $\theta_{\alpha}$ and $\theta_s$ are the phases of $\alpha$ and $s$, respectively, and where we have defined the following dimensionless parameters:
\begin{equation*}
	x \equiv \frac{\vert s \vert}{\vert \tilde{s_c} \vert}, \quad s_c \equiv \frac{\vert \tilde{s_c} \vert}{m_P}, \quad m \equiv \frac{M}{m_P}.
\end{equation*}
The symbol $\vert \tilde{s_c} \vert$, as before, denotes the inflaton field at the waterfall point and its dimensionless value $s_c$ is given by
\begin{equation}\label{critical}
	-\kappa m^2 + {s_c}^2 (\kappa + 4 |\alpha| s_c) = 0.
\end{equation}
It can be shown that, using just Equation \eqref{VGlobal}, one cannot obtain a red-tilted spectrum while simultaneously satisfying the slow-roll conditions (to be defined somewhat later). An analytical calculation reveals that, by imposing the condition $n_s < 1$ in the slow-roll approximation, the inflaton field at the start of inflation is necessarily trans-Planckian (see Appendix \ref{app2}). We therefore cannot, in this scenario, achieve a suitable spectral tilt without additional terms. 

In order that our model yield more experimentally favorable results, we include soft \cite{csenouguz2005reheat} and SUGRA corrections \cite{linde1997hybrid, csenouguz2005reheat} to the global plus RC potential. The soft terms are derived in a gravity-mediated SUSY-breaking scenario \cite{ *[{}] [{; References to the original articles may be found within.}] nilles1984supersymmetry}; including the soft mass squared terms, they are
\begin{equation*}
	a\, m_{3/2} \,\kappa m^2\,{s_c} x \quad , \quad {m_{3/2}}^2\, {s_c}^2 x^2 \quad , \quad b\, m_{3/2}\, \vert \alpha \vert \,{s_c}^4 x^4,
\end{equation*}
where the last term is a direct consequence of our $S^4$ term in $W$, and $m_{3/2}$ is the gravitino mass ($\sim$ TeV) divided by the Planck scale.  We write the effective coefficients of the soft terms as
\begin{equation}\label{softcoefficients}
	\begin{split}
		a &= 2 \left[ 2 \cos(\theta_s ) - \vert A \vert \cos (\theta_A + \theta_s) \right], \\
		b &= 2 \left[ \vert A \vert \cos(\theta_A + \theta_\alpha + 4\theta_s) +  \cos (\theta_\alpha + 4 \theta_s) \right],
	\end{split}
\end{equation}
where each $\theta_i$, $i \in \{A, \alpha, s\}$, is the phase of a complex parameter, and $A$ is the trilinear coupling. Note that we cannot take $\alpha$ real without loss of generality, since we have already absorbed the phase of $\kappa$ into that of $s$; therefore, we consider the most general case where $\alpha$, $s$, and $A$ are complex.

While $\theta_A$ and $\theta_\alpha$ are components of couplings, $\theta_s $ is a dynamical field. For the sake of simplicity, we minimize the potential with respect to $\theta_s $ so as to define the inflaton field purely as $\vert s \vert$. As a result (see Appendix \ref{app1}), we choose the following values of the phases 
\begin{equation*}
	\theta_s = l \pi \quad,\quad \theta_A = n \pi \quad,\quad \theta_\alpha = p \pi,
\end{equation*}
such that $l$, $n$, and $p$ are all odd integers. With these choices, the effective coefficients are
\begin{equation*}
	a = -2 (2 + \vert A \vert ) \quad, \quad b = 2 (\vert A \vert - 1),
\end{equation*}
and in conjunction with these phase choices we additionally impose the condition that $\vert A \vert <1 $, or equivalently $b < 0 $ (see Appendix \ref{app1}). Henceforth, we drop the bars on $A$ and $\alpha$ with the understanding that they represent the moduli of the corresponding complex quantities.

The SUGRA scalar potential is given by
\begin{equation}\label{Vlocal}
	V_{F\text{;SUGRA}} = e^{K/{m_P}^2} \left( K_{ij}^{-1}D_{z_i}WD_{z^*_j}W^* - 3{m_P}^{-2}\left| W\right|^2 \right),
\end{equation}
with 
\begin{equation*}
	K_{ij} \equiv \frac{\partial ^2 K}{\partial z_i\partial z_j^*} \quad , \quad D_{z_i}W \equiv \frac{\partial W}{\partial z_i}+{m_P}^{-2}\frac{\partial K}{\partial z_i}W \quad , \quad D_{z_i^*}W^*=\left(D_{z_i}W\right)^*.
\end{equation*}
Note that Equation \eqref{Vlocal} reduces to Equation \eqref{VFglobalgeneral} in the limit $m_P \rightarrow \infty$ and where minimal K\"ahler is used. (Minimal K\"ahler is defined as ${\vert S \vert}^2 + {\vert \overline{\Phi} \vert}^2 + {\vert \Phi \vert}^2$.)

We include SUGRA correction terms up to sixth order in the inflaton field $\vert s \vert$, consistent with our inclusion of the $\alpha^2 {\vert s \vert}^6$ global SUSY term; they are:
\begin{equation*}
	\frac{1}{2} \kappa^2 m^4 {s_c}^4 {x}^4 \quad, \quad \frac{2}{3} m^4 {\kappa}^2 {s_c}^6 x^6 \quad, \quad -12 \kappa \, m^2 \,\alpha \, {s_c}^5 x^5.
\end{equation*}
Hence, with the addition of the soft SUSY-breaking, SUGRA, and 1-loop radiative correction terms to Equation \eqref{VGlobal}, the full scalar potential, scaled by $1/{{m_P}^4}$, becomes:
\begin{equation}\label{Vfinal}
\begin{split}
		\mathcal{V} &= \kappa^2 m^4 - 8 \alpha \, \kappa \, {s_c}^3 m^2 x^3 + 16 \, \alpha^2 \, {s_c}^6 x^6 + \frac{m^4 \kappa^4\mathcal{N}}{8\pi^2}F(x) \\ 
		&+ a\, m_{3/2} \,m^2 \,\kappa \,s_c\,x + b \,m_{3/2} \, \alpha \, {s_c}^4 x^4 + {m_{3/2}}^2 \, {s_c}^2  x^2\\ 
		&+ \frac{1}{2} m^4 \kappa^2 {s_c}^4 x^4 + \frac{2}{3} m^4 \kappa^2 {s_c}^6 x^6  -12 \kappa \, m^2 \,\alpha \, {s_c}^5 x^5. 
\end{split}
\end{equation}

In solving the essential cosmological equations, we employ the slow-roll approximation throughout, in which inflation occurs while the slow-roll parameters are less than unity. In our notation these are written as:
\begin{equation*}
	\epsilon = \frac{1}{4 {s_c}^2} {\left(\frac{\mathcal{V}'}{\mathcal{V}} \right)}^2 \quad , \quad \eta = \frac{1}{2 {s_c}^2} \frac{\mathcal{V}''}{\mathcal{V}} \quad , \quad \xi^2 = \frac{1}{4 {s_c}^2} {\frac{\mathcal{V}''' \mathcal{V}'}{\mathcal{V}^2}}.
\end{equation*}
Here, the prime ( $'$ ) denotes a derivative with respect to $x$. Inflation ends either when the slow-roll parameters become unity, or when the inflaton field reaches the waterfall point at $x = 1$. Observable inflation starts at $x_0$, defined at the pivot scale $k_0 = 0.002\,\text{ {Mpc}}^{-1}$, and ends at $x_{e}$. With this, the number of e-foldings becomes, to leading order,
\begin{equation}\label{N_0}
	N_0 \approx 2 {s_c}^2 \, \int_{x_e}^{x_0} \frac{\mathcal{V}}{\mathcal{V}'}\,dx,
\end{equation}
while the usual definitions hold for
\begin{equation}\label{r_ns}
	r \approx 16 \epsilon \quad , \quad n_s \approx 1 - 6 \epsilon + 2 \eta \quad , \quad \frac{\text{d} n_s}{\text{d} \ln k} \approx 16 \epsilon\, \eta - 24 \epsilon^2 - 2 \xi^2.
\end{equation}
The amplitude of the curvature perturbation is given, to leading order, by
\begin{equation}\label{deltaRsq}
	\Delta^2_\mathcal{R} \approx  \frac{{s_c}^2}{6 \pi^2} \, \frac{\mathcal{V}^3}{\mathcal{V}'^2}.
\end{equation}
For higher order expressions see \cite{csenouguz2008chaotic}. Note that Equations \eqref{r_ns} and \eqref{deltaRsq} are evaluated at the pivot scale.

In our numerical calculations we take $m_{3/2} = 1\text{ TeV}/m_P$, $Q = 10^{15}\text{ GeV}/m_P$, and since we are implicitly embedding our model in flipped $SU(5)$, we take $\mathcal{N} = 10$ \cite{rehman2010minimal}. In addition, we impose the ranges in Table \ref{rangetable}.
\begin{table}[Ht]
	\begin{tabular}{c|c|c||c|c}
		{\bf Fundamental} & {\bf Range} & {\bf Scale} & {\bf Derived} & {\bf Constraining range} \\
		{\bf parameter} &  & {\bf type} & {\bf quantity} &  \\
		\hline
		$\kappa$ & $[10^{-6}, 1]$ & log & $n_s$ & $[0.92, 1.02]$ \\
		$m$ & $[10^{-4}, 10^{-1}]$ & log &  & $ \approx 0.971 \pm 4\sigma$ \\ \cline{4-5}
		$\alpha$ & $[10^{-14}, 10^{-8}]$ & log & $\Delta_\mathcal{R}^2$ & $[2.271, 2.583] \times 10^{-9}$ \\ 
		$A$ & $[10^{-10}, 1]$ & log &  & $= 2.427 \times 10^{-9} \pm 2\sigma$ \\
		\cline{4-5}
		$x$ & $(1, \frac{1}{m}]$ & linear & $r$ & $< 0.13$ \\
		$s_c$ & $(0, 1)$ & linear & $N_0$ & $[50,60]$
	\end{tabular}
	\caption{These are the ranges specified for the fundamental parameters in Equation \eqref{Vfinal}, and constraints placed on derived quantities, that we have used in our numerical calculations. Note that $x$ can take on any value between the waterfall point and the Planck scale. The experimental bounds on $r$, $\Delta_\mathcal{R}^2$, and $n_s$ are from the WMAP nine-year analysis + eCMB + BAO + $H_0$ \cite{WMAPnine}. The numerical constraints on the quantities $r$ and $\Delta_\mathcal{R}^2$ are the experimental bounds; however, the numerical constraints on $n_s$ differ slightly from the experimental bounds. This has been done for ease of plotting.}
	\label{rangetable}
\end{table}

\section{Results}\label{sec_Results}
\subsection{Overview}
\begin{figure}[h]
	\subfigure[]{
	\includegraphics[scale=0.245]{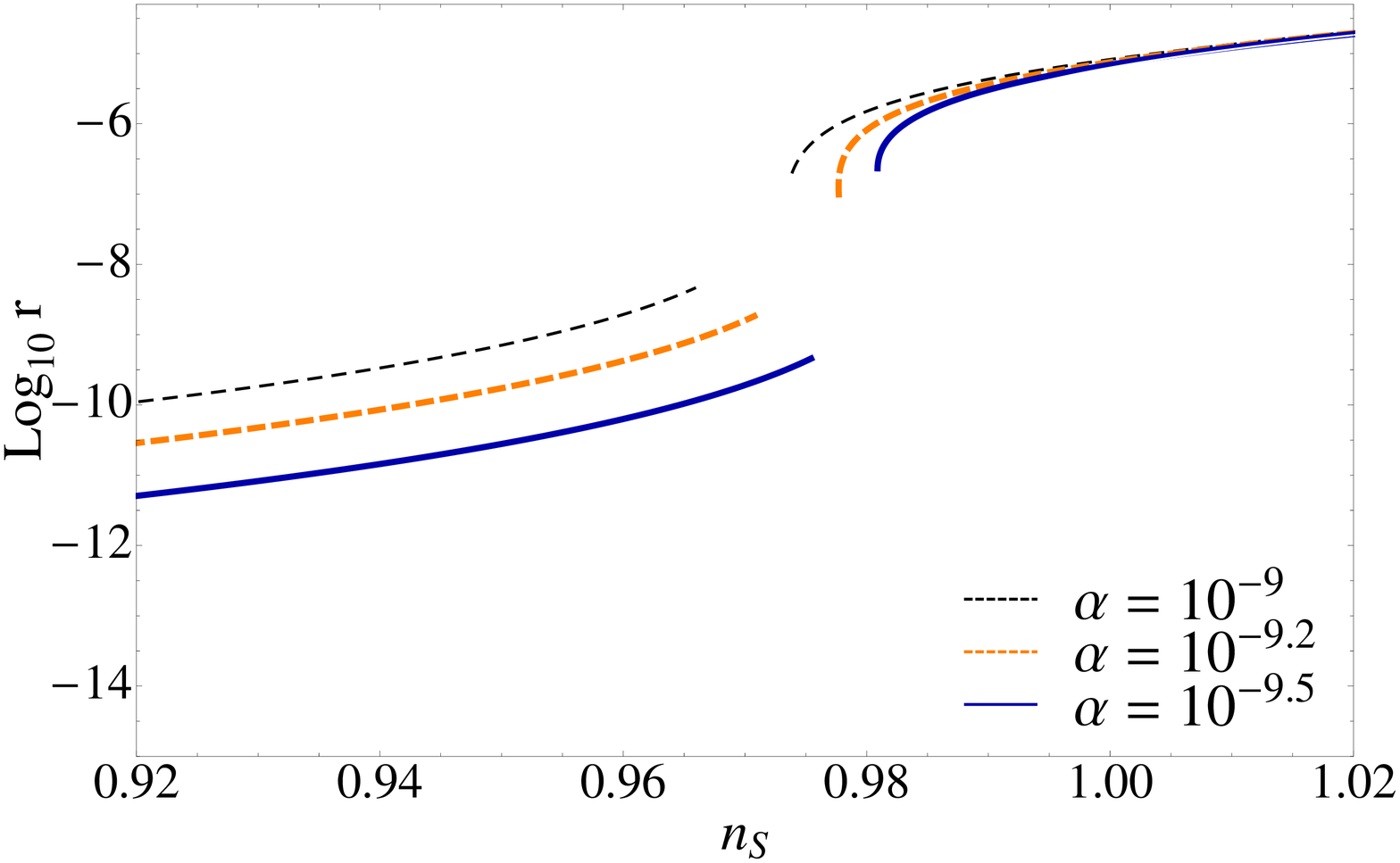}\label{larger}}
	\subfigure[]{
	\includegraphics[scale=0.245]{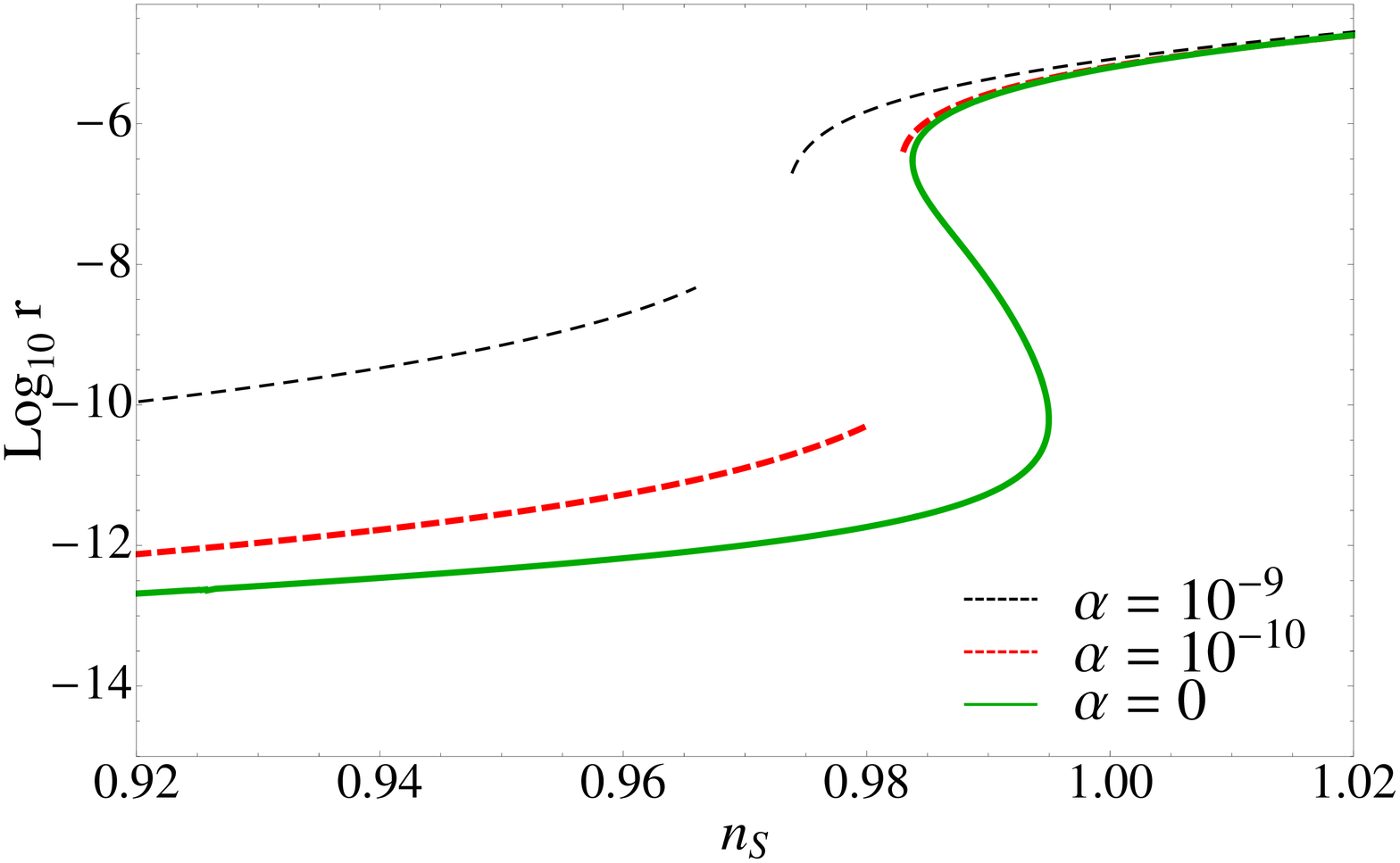}\label{fuller}}
	\caption{The tensor-to-scalar ratio, $r$, versus the scalar spectral index, $n_s$, is depicted. Three curves for large values of $\alpha$ are shown in \ref{larger}, and a larger range of $\alpha$ is taken in \ref{fuller}. Here, the number of e-foldings and $A$ have been fixed at $50$ and $10^{-4}$, respectively. With $\alpha = 0$, we produce solutions closely matching the $a = -1$ case in \cite{rehman2010supersymmetric}. Note that this curve (in \ref{fuller}) does not produce false vacua.}
	\label{r_vs_ns}
\end{figure}

Previous studies have shown that small gravity waves are generated using minimal K\"ahler and a TeV-scale positive soft SUSY-breaking mass squared term (i.e., ${m_{3/2}}^2 \, x^2$, with $m_{3/2} \sim 10^{-16}$) \cite{rehman2010supersymmetric, civiletti2011red, rehman2010minimal}. Specifically, when the lowest-order SUGRA correction term and a negative linear soft term ($a = -1$) are added to the global SUSY plus TeV-scale positive soft mass squared plus RC potential, one finds that $r \sim 10^{-12.5}$ around the WMAP nine-year central value $n_s = 0.971$ \cite{rehman2010supersymmetric}. Alternatively, using non-minimal K\"ahler in shifted inflation with positive TeV-scale soft mass squared and $a = 1$, $0$, or $-1$ \cite{civiletti2011red} (See also \cite{rehman2010observable, rehman2012simplified, ur2007supersymmetric} for further references.), or non-minimal K\"ahler with with same $a$ values, and large, positive soft mass squared terms ($\sim 10^{-5}$) \cite{shafi2011observable}, one can generate $r \sim 10^{-2}$ with red spectral tilt. In this paper we find that the solutions follow curves of a similar shape to those presented in \cite{rehman2010supersymmetric}, as can be seen in Figure \ref{r_vs_ns}.
By employing minimal K\"ahler, positive TeV-scale soft mass squared terms, and a negative linear and a negative $\alpha$-dependent quartic soft term, we obtain in this paper $r \sim 10^{-8.5}$ around $n_s \simeq 0.965$ for $\alpha = 10^{-9}$; in \cite{rehman2010supersymmetric}, one obtains $r$ values four orders of magnitude lower than this, at $n_s \simeq 0.965$. As we will describe, this model yields even larger gravity waves ($\sim 10^{-4}$, see Figure \ref{fullrns}) with red spectral tilt, and we expect that with non-minimal K\"ahler this model can yield solutions similar to \cite{shafi2011observable}. While the full set of results is outside the reach of current experiments such as Planck, a model in which solutions are tending toward larger $r$ solutions is nonetheless preferred.

\subsection{The Effect of the Parameters on the Model}
Our potential is dependent upon the inflaton field $x$ and the parameters $A$, $\alpha$, $\kappa$, and $m$. Our new parameter $\alpha$, which parametrizes the amount of $R$-symmetry violation beyond the renormalizable level, yields qualitatively and quantitatively distinct results from the standard case. 

\begin{figure}[h]
	\subfigure[]{
		\includegraphics[scale=0.85]{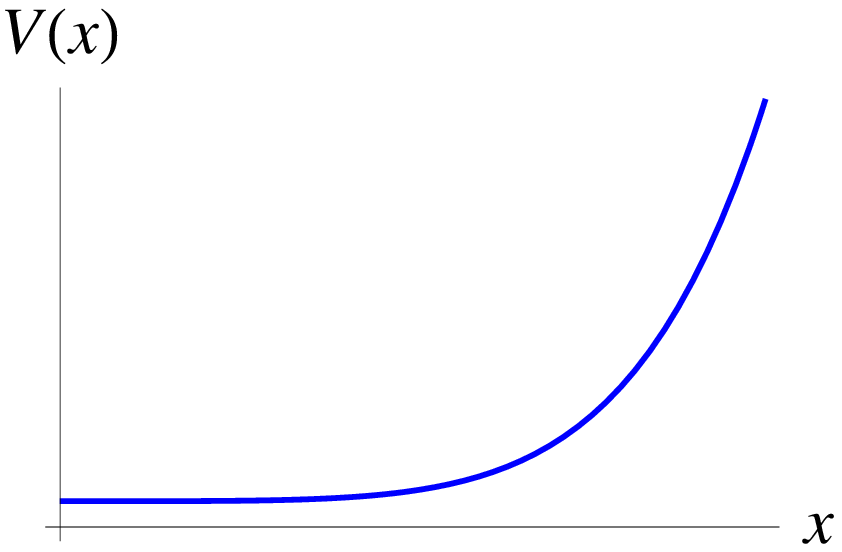}\label{V1}}
	\subfigure[]{
		\includegraphics[scale=0.85]{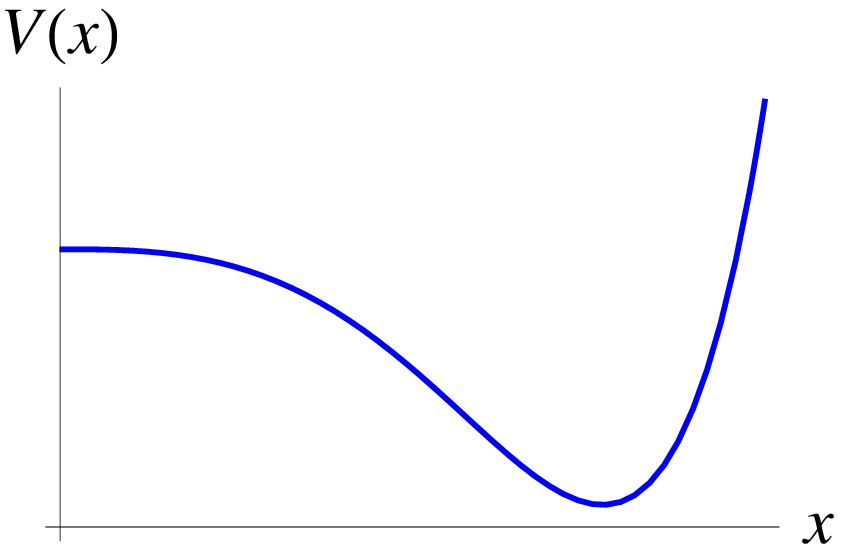}\label{V2}}
	\caption{The qualitative change in behavior caused by the negative $\alpha$ terms. Figure (a) depicts a potential that is well-behaved, i.e., the field will roll toward the global minimum, while Figure (b) depicts a potential with a false vacuum. Mathematical solutions which produce false vacua are not acceptable inflationary scenarios.}
\end{figure}

The negative $\alpha$-dependent terms in \eqref{Vfinal} create false vacua in some regions of parameter space, i.e. the general behavior of the potential changes from that of Figure \ref{V1} to that of \ref{V2}.
(Note that the inflaton field rolls from right to left in these figures.) We cannot produce a successful inflationary scenario from Figure \ref{V2} as the system will become trapped in the false vacuum. Rejecting solutions for which this occurs produces gaps in the parameter space such as those seen in Figures \ref{r_vs_ns} and \ref{figpanel_NN2}. Note that these vacua did not appear in \cite{rehman2010supersymmetric}.

Figure \ref{r_vs_ns} depicts the effects of $\alpha$ on $r$. (See Figure \ref{figpanel_NN2} for further results. Note the similarity of these curves to those in \cite{rehman2010supersymmetric}.) The potential in Equation \eqref{Vfinal} differs from that in \cite{rehman2010supersymmetric} by two higher-order SUGRA correction terms, an $\alpha$-dependent quartic soft term, and two global $\alpha$-dependent terms ($a$ is also different). The effect of $\alpha$ is to raise $r$, particularly in red-tilted regions. This is primarily a result of the global term proportional to $\alpha$. The $\alpha =  10^{-9}$ curve is raised by three to six orders of magnitude for $0.92 <  n_s < 0.98$, as compared to the $\alpha = 0$ case in \cite{rehman2010supersymmetric}.
 
Our model greatly benefits from the fact that we can, as noted, generate larger gravity waves than the standard ($\alpha = 0 $) case. However, $\alpha$ cannot be raised arbitrarily, since we require $R$-symmetry violation to be small. We find no need to impose an upper bound, though, because our parameter study yields a numerical upper bound $\alpha \sim 10^{-7}$. This can be understood mathematically by noting that $\vert \mathcal{V}' \vert$ increases faster than $\vert \mathcal{V} \vert$ as  $\alpha \rightarrow 10^{-7}$ (from smaller $\alpha$). The e-foldings constraint \eqref{N_0} becomes impossible to satisfy at large $\alpha$, because its integrand is suppressed by a large $\mathcal{V}'$, and the limits of the integral can only be marginally altered.

\begin{figure}[Ht]
	\subfigure[]{  
		\includegraphics[trim = 200 0 25 25, scale=0.36]{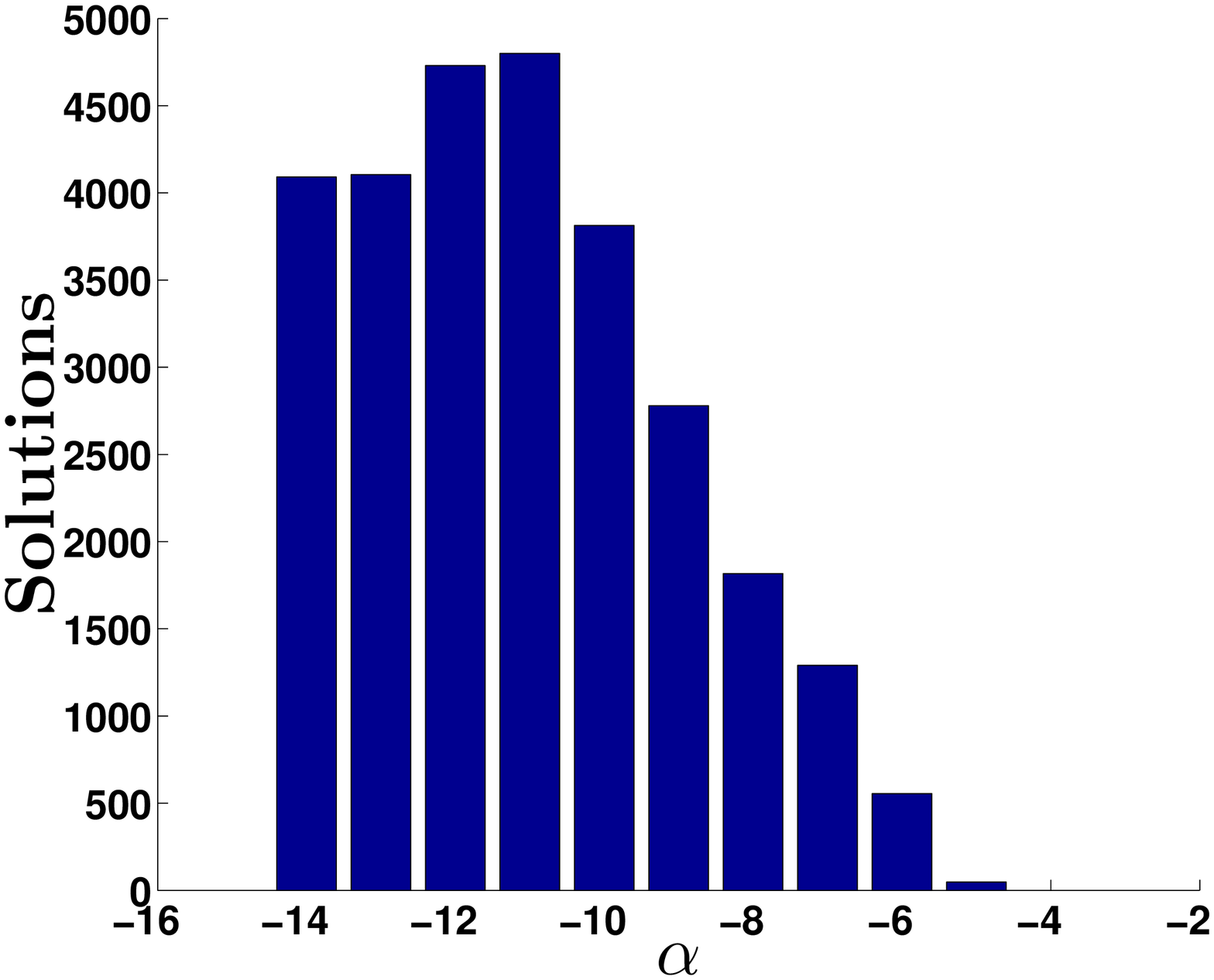}\label{hist1}}
	\subfigure[]{
		\includegraphics[trim = 55 0 200 25, scale=0.36]{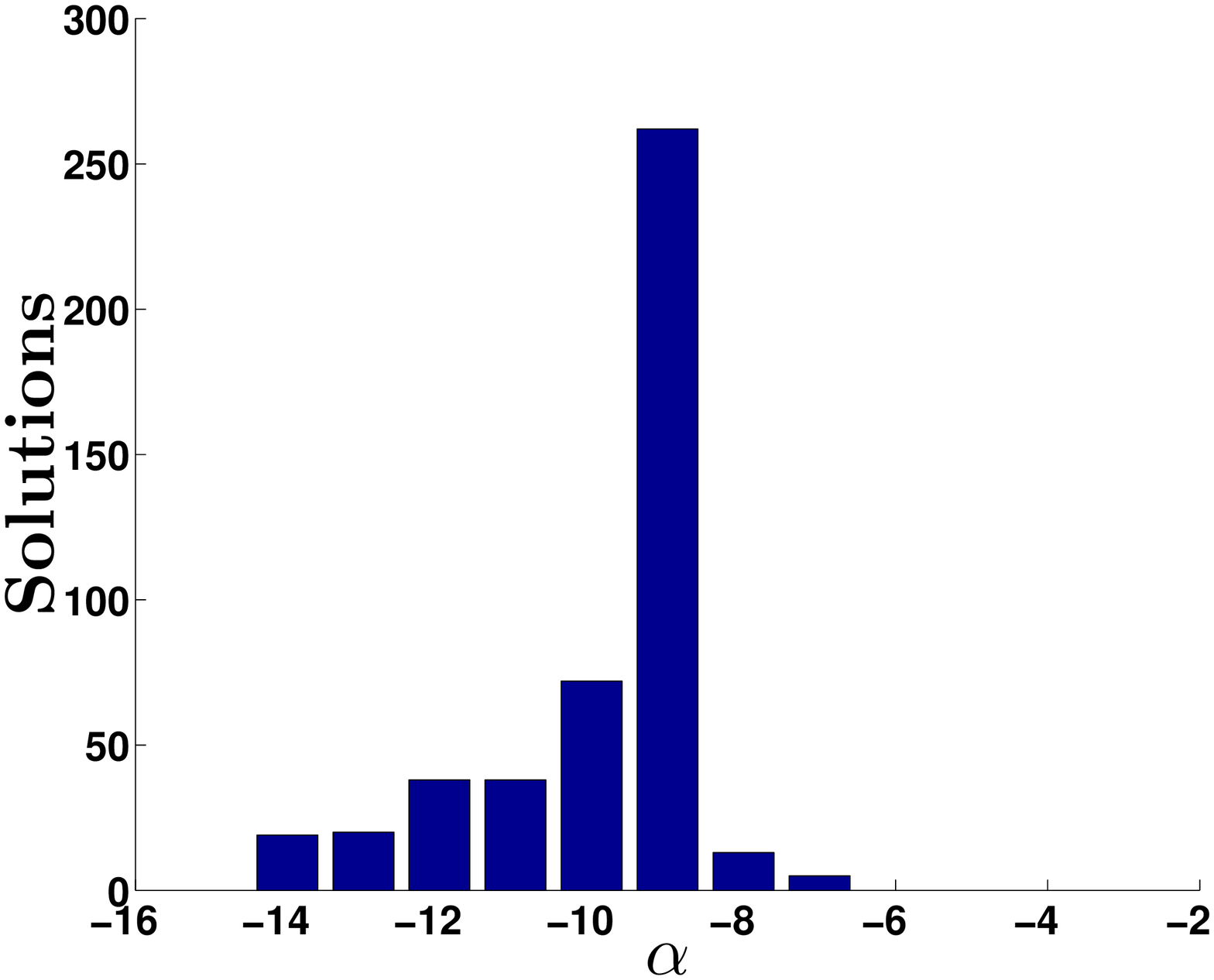}\label{histN0}}
	\caption{The number of solutions to $\Delta_\mathcal{R}^2 = 2.427 \times 10^{-9}$ (a), and both $\Delta_\mathcal{R}^2 = 2.427 \times 10^{-9}$ and $N_0 \in [50,60]$ (b).}
\end{figure}

We find that $x_0$ can vary over at least two orders of magnitude until $\alpha \sim 10^{-8}$; then, $x_0$ is compressed to $\sim 10$. Likewise, the end of inflation is pushed toward waterfall, $x_{e} = $ 1, as $\alpha \rightarrow 10^{-7}$. Thus the distance in $x$ over which inflation occurs approaches an approximately constant value.  If we take solutions to $\Delta_\mathcal{R}^2 = 2.427 \times 10^{-9}$ (Figure \ref{hist1}) and then impose the constraint that the number of e-foldings be between 50 and 60 (Figure \ref{histN0}), we observe that, for many orders of magnitude in $\alpha$, requiring sufficient inflation decreases by at least an order of magnitude the number of solutions relative to those obtained merely from the curvature perturbation constraint.
Achieving a sufficient amount of inflation severely limits the number of viable solutions generally, but is most limiting at large $\alpha$. The curvature perturbation constraint also has a limiting effect as $\alpha$ increases, although notice that without ensuring a sufficient amount of observable inflation, one can obtain ``solutions" up to $\alpha = 10^{-5}$ (Figure \ref{hist1}).

\begin{figure}[Ht]
	\subfigure[]{  
		\includegraphics[trim = 200 0 10 25, scale=0.36]{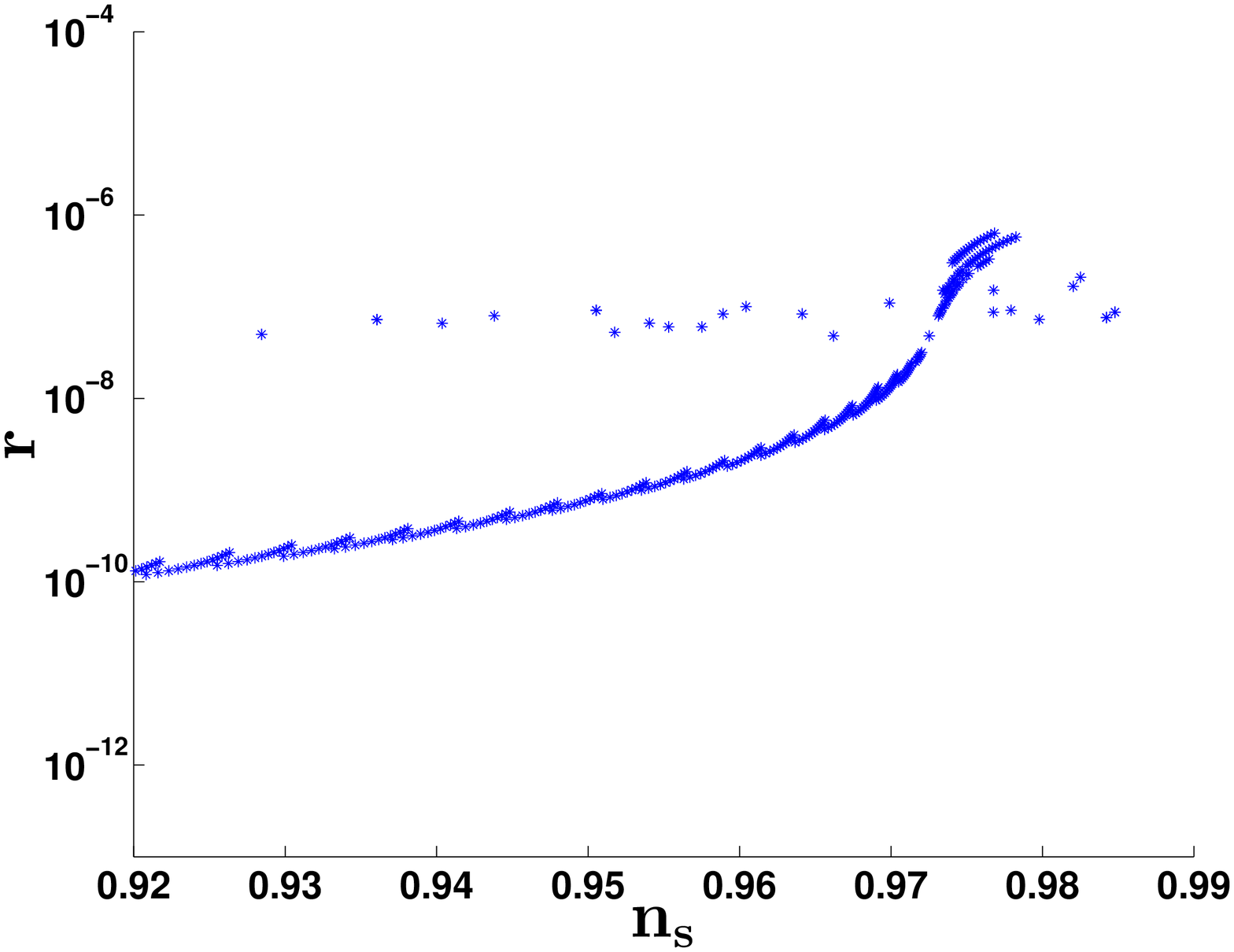}\label{r_vs_nsFIXED}}
	\subfigure[]{ 
		\includegraphics[trim = 70 0 200 25, scale=0.36]{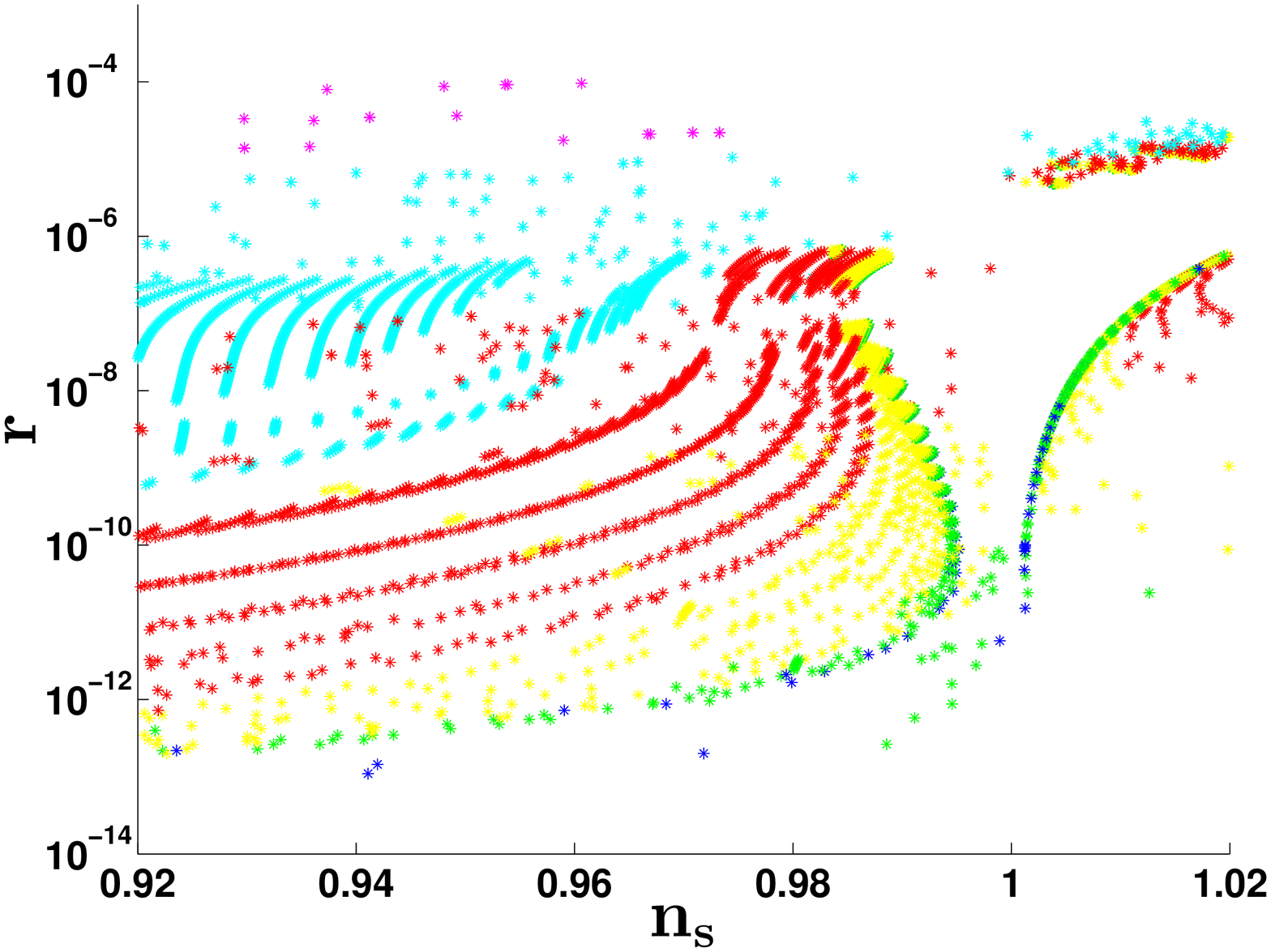}\label{fullrns}}
	\caption{(a) The tensor-to-scalar ratio, $r$, versus the scalar spectral index, $n_s$, for $\alpha = 10^{-9}$, $A = 10^{-4}$, and $N_0 \in [50,60]$. One can see the ``horizontal solutions", which yield $r$ values somewhat above $10^{-8}$. (b) The tensor-to-scalar ratio, $r$, versus the scalar spectral index, $n_s$ for solutions corresponding to the ranges in Table \ref{rangetable}. Solutions are color-coded as follows: blue - ($10^{-14} < \alpha \leq 10^{-12}$), green - ($10^{-12} < \alpha \leq 10^{-11}$), yellow - ($10^{-11} < \alpha \leq 10^{-10}$), red - ($10^{-10} < \alpha \leq 10^{-9}$), cyan - ($10^{-9} < \alpha \leq 10^{-8}$), magenta - ($10^{-8} < \alpha \leq 10^{-7}$).}
	\label{fig5}
\end{figure}

\begin{figure}[Ht]
	\subfigure[]{  
		\includegraphics[trim = 200 0 10 25, scale=0.36]{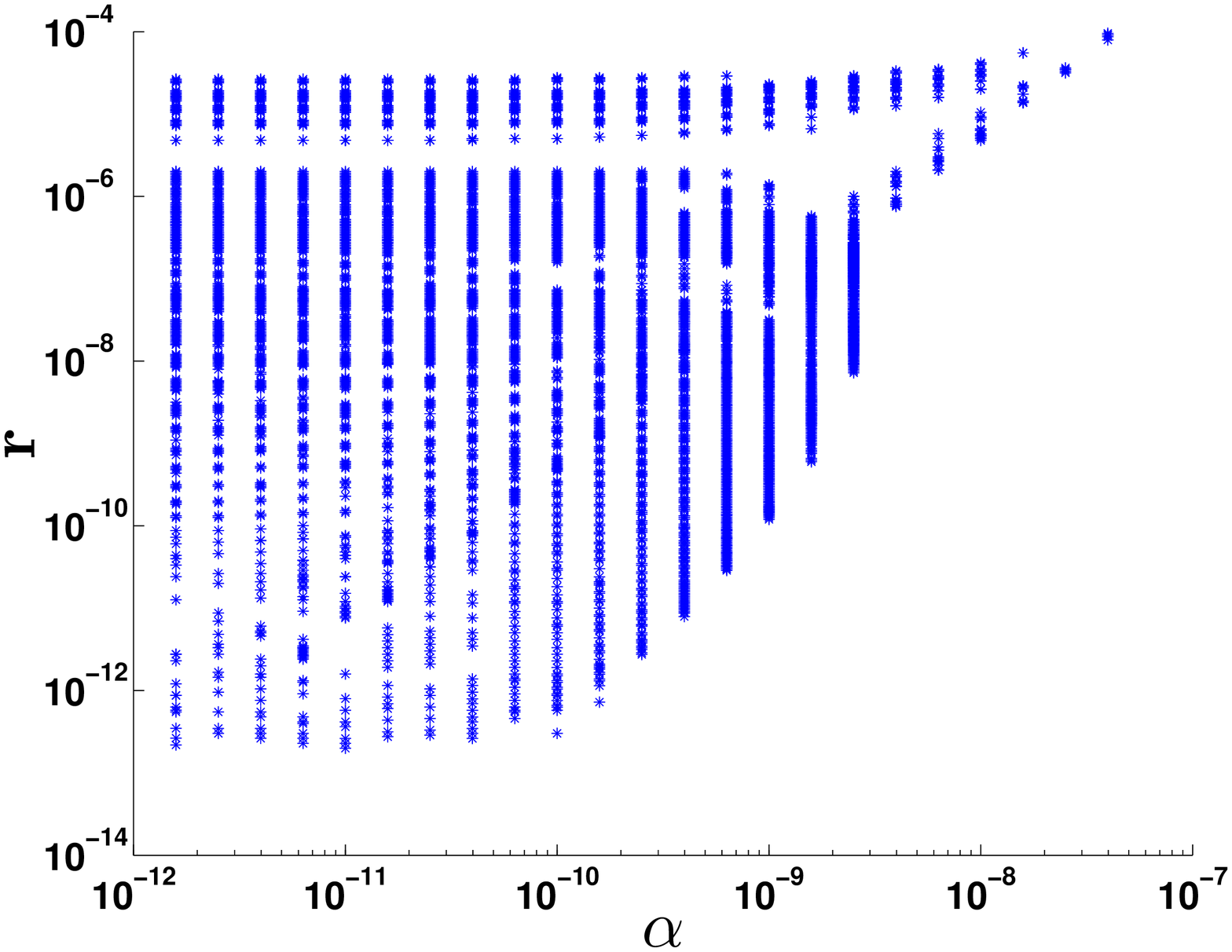}\label{r_vs_alpha}}
	\subfigure[]{ 
		\includegraphics[trim = 60 0 200 25, scale=0.36]{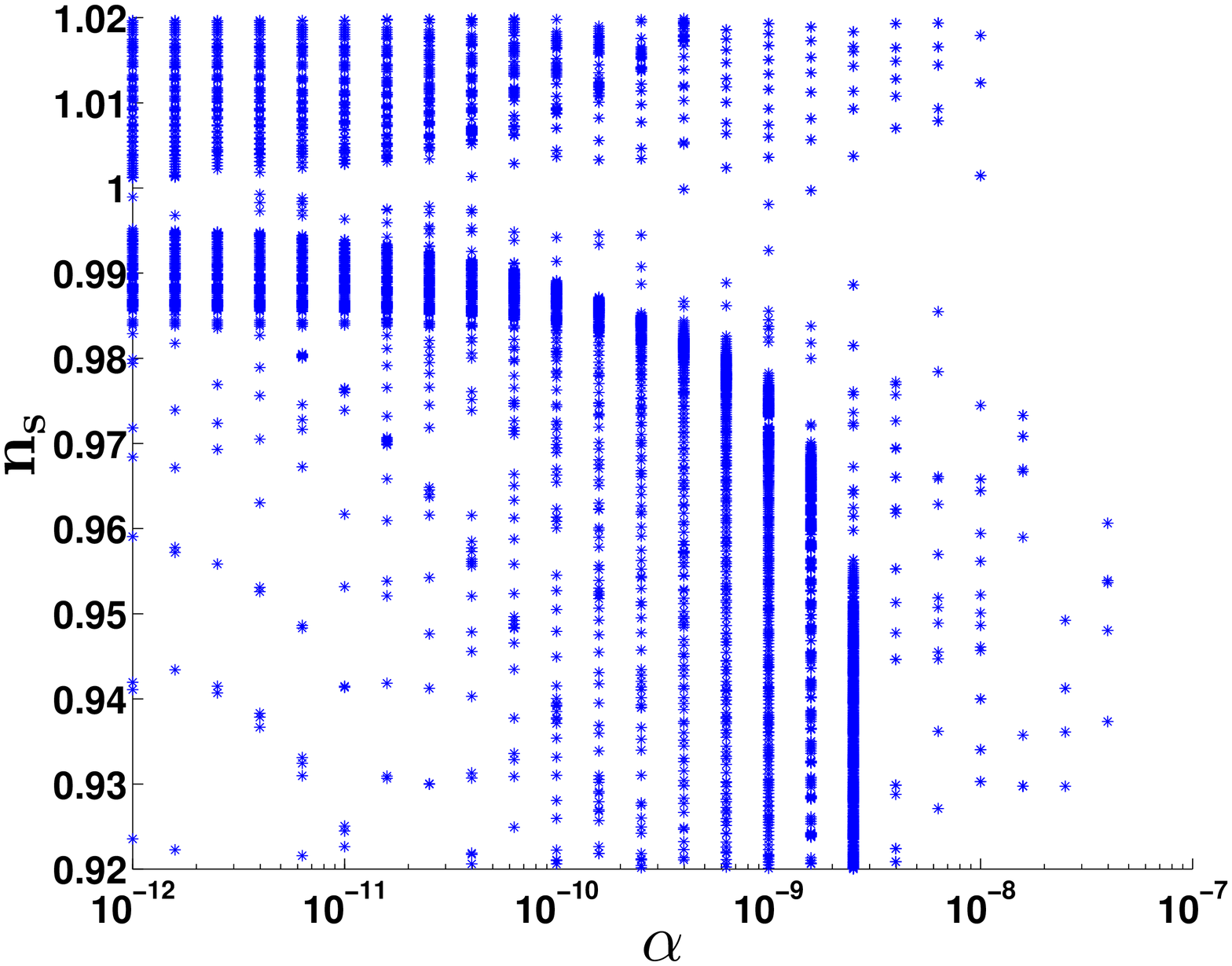}\label{ns_vs_alpha}}
	\caption{(a) The tensor-to-scalar ratio, $r$, versus $\alpha$ using the constraints in Table \ref{rangetable}. (b) The scalar spectral index, $n_s$, versus $\alpha$ using the constraints in Table \ref{rangetable}. Note that in the red-tilted region where $\alpha \rightarrow 0$, solutions are sparse.}
	\label{fig6}
\end{figure}

\begin{figure}[ht]
	\subfigure[]{
		\includegraphics[scale=0.275]{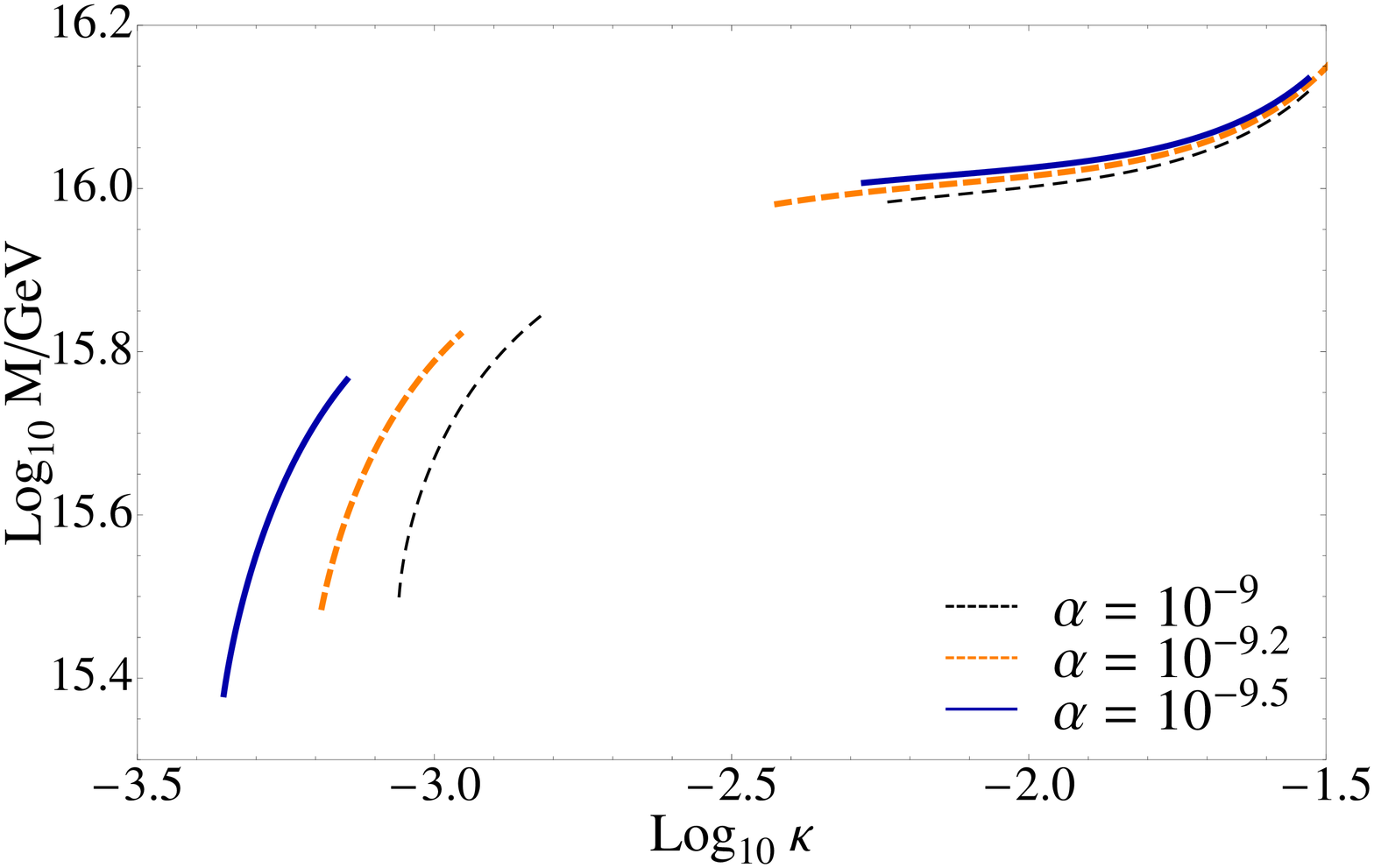}\label{M_vs_kappa}}
	\subfigure[]{
		\includegraphics[scale=0.275]{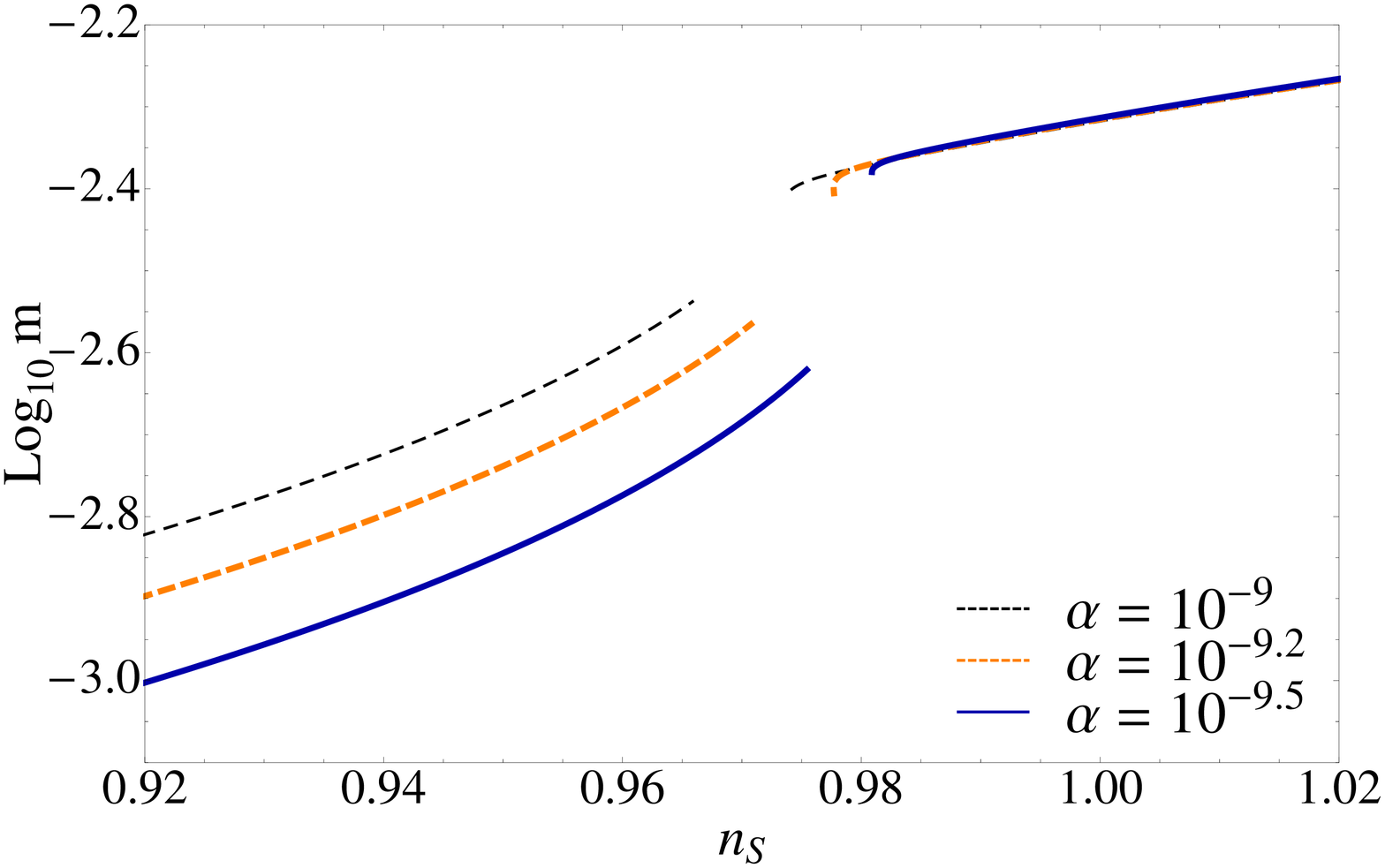}\label{m_vs_ns}}
	\subfigure[]{
		\includegraphics[scale=0.275]{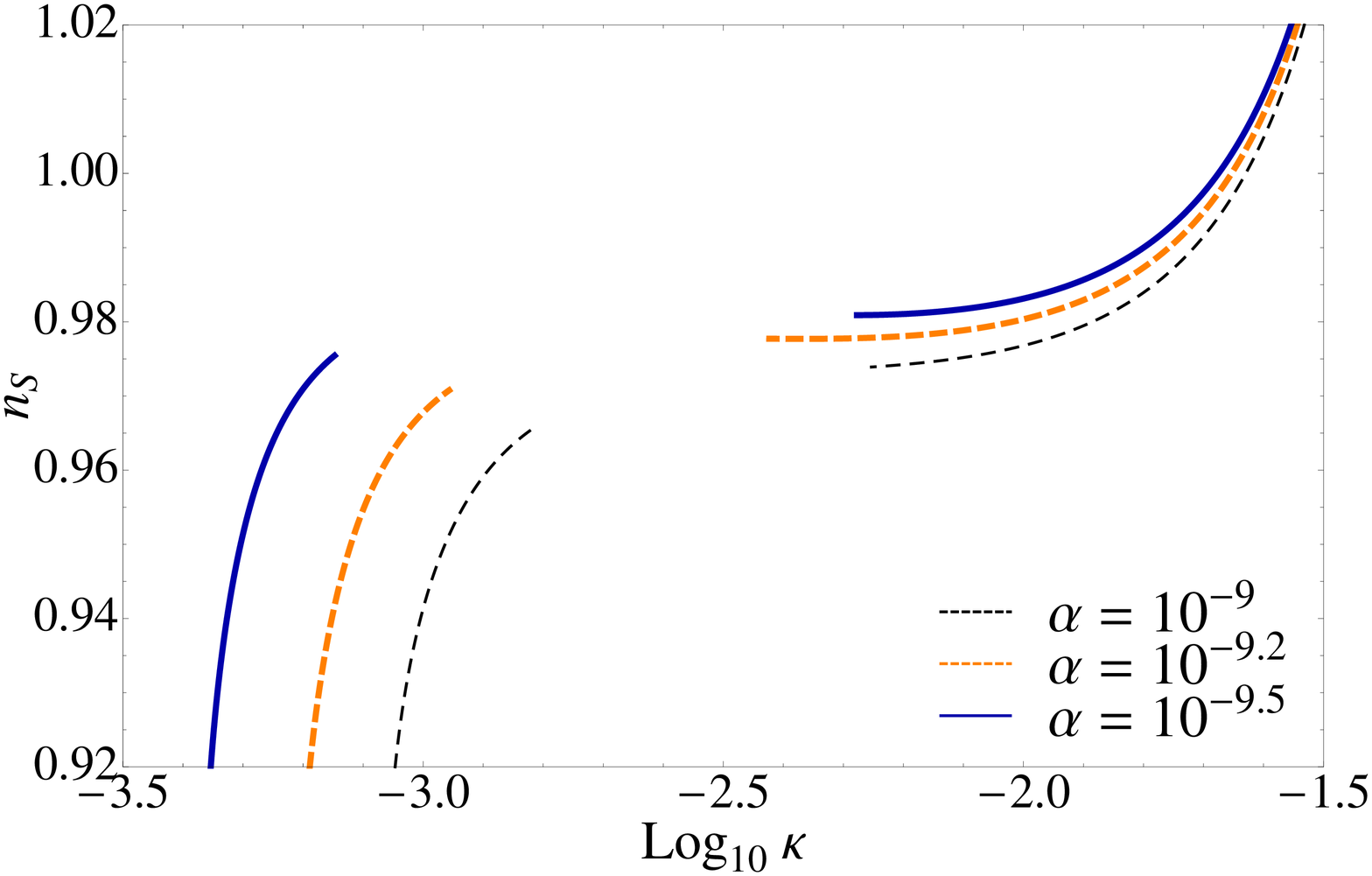}\label{ns_vs_kappa}}
	\subfigure[]{
		\includegraphics[scale=0.275]{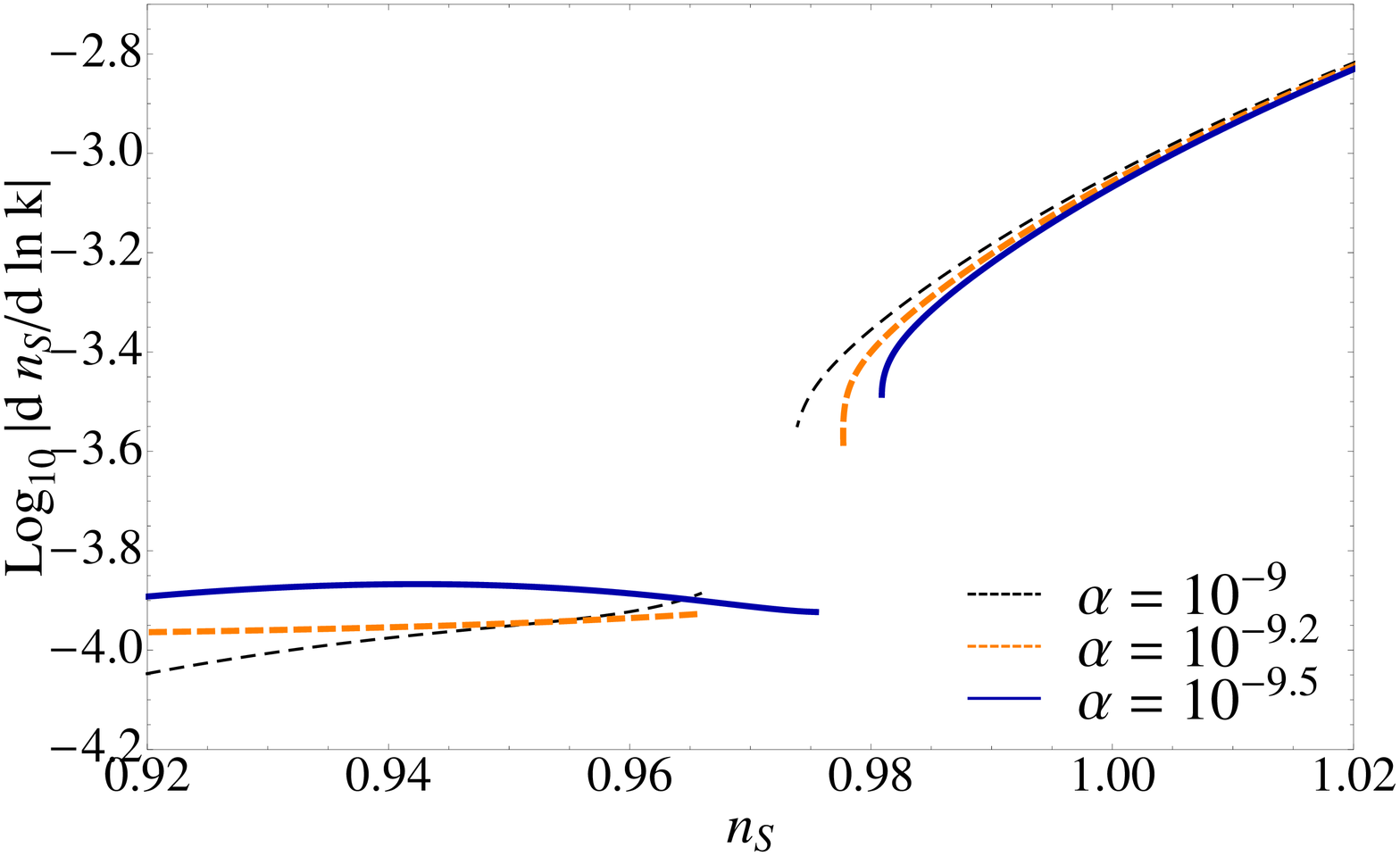}\label{runFig}}
	\caption{Here we plot our numerical results in which the number of e-foldings and $A$ have been kept fixed at $50$ and $10^{-4}$, respectively. The absolute value of the running of $n_s$, $\log_{10} \vert \frac{\text{d} n_s}{\text{d} \ln k} \vert$, is plotted in \ref{runFig}.}
	\label{figpanel_NN2}
\end{figure}

Numerically, we obtained our results using two independent methods: a continuation method and a parameter study. The results of the former are seen in Figures \ref{r_vs_ns} and \ref{figpanel_NN2}. The latter results are presented in Figures \ref{fig5} and \ref{fig6}. Figure \ref{r_vs_nsFIXED} shows the presence of qualitatively new ``horizontal solutions" in the $r-n_{s}$ plane. If we drop all the $\alpha$-dependent terms in the scalar potential \eqref{Vfinal} except the global term linear in $\alpha$ (which is $- 8 \alpha \, \kappa \, {s_c}^3 m^2 x^3$), the number of horizontal solutions in this region increases. On the other hand, dropping this term and keeping the others does not produce a viable inflationary scenario. Since the global $\alpha$ term is the most dominant $\alpha$-dependent term, it is primarily responsible for producing these horizontal solutions.

We find that limits on $r$ naturally arise in our model (see Figure \ref{r_vs_alpha}). Solutions producing $r \sim 10^{-4}$ can be produced throughout the range of $\alpha$ that we have taken, although all of these except the solutions near the upper limit of $\alpha$ correspond to a blue-tilted spectrum. Figure \ref{fullrns} depicts this behavior; note that the only large-$r$ solutions  corresponding to red spectral tilt are colored magenta and cyan (meaning that $10^{-9} < r \leq 10^{-7}$). Figure \ref{r_vs_alpha} also indicates that the smallest-$r$ solutions that can be produced are increasingly larger as $\alpha$ increases, so that at the upper limit of $\alpha$ only $r \sim 10^{-4}$ can be produced. This effect can be understood by using the following approximation of the energy scale of inflation: ${\mathcal{V}_0}^{1/4} \sim \kappa^{3/2} m^2/\sqrt{\lambda^2 {x_0}^2 \alpha}$, where $\lambda \equiv [(3456 \, \Delta_\mathcal{R}^2\big \vert_{x_0} \pi^2)]^{1/4}$. As mentioned, the integrand in Equation \eqref{N_0} is suppressed by the fact that $\vert \mathcal{V}' \vert$ increases faster than $\vert \mathcal{V} \vert$ as $\alpha$ increases. To compensate for this effect the $\kappa^2 m^4$ term in $\mathcal{V}$ increases, raising the numerator in ${\mathcal{V}_0}^{1/4}$. This prohibits small-$r$ solutions in large-$\alpha$ regions (recall that $r \propto {\mathcal{V}_0}^{1/4}$).

The upper limit on $r$ of $\sim 10^{-4}$, approximately constant over many orders of magnitude of $\alpha$, is a consequence of the Lyth Bound \cite{lyth1997would} $r \lesssim \mathcal{O}(10^{-2}) \times m^2 (x_0 - x_e)^2$. The largest $m$ values obtainable ($m \sim 10^{-1}$) correspond to very small values of $x_0 - x_e$, and therefore limit $r$ below $10^{-4}$. Solutions with $m \sim 10^{-2}$ correspond to $x_0 \approx 10$ and $x_e \approx 1$; thus, $m^2 (x_0 - x_e)^2 \sim 10^{-2}$ for the largest $r$ values obtainable in our model.

The parameter $A$, arising from the gravity-mediated soft-SUSY breaking terms, does not have any discernible effect on our results in the red-tilted region. This is expected from the fact that the soft terms are suppressed by the gravitino mass ($\sim$ TeV).

The plots in Figure \ref{figpanel_NN2} depict the effects of increasing $\alpha$. Figures \ref{M_vs_kappa}, \ref{ns_vs_kappa}, and \ref{runFig} are in direct reference to \cite{rehman2010supersymmetric}. We can see from Figure \ref{ns_vs_kappa} that large $\alpha$ boosts $\kappa$, especially in the red-tilted region. Similarly, Figure \ref{M_vs_kappa} shows that, for the same breaking scale $M$, $\kappa$ is boosted as $\alpha$ increases. Note that $m$ is also boosted by $\alpha$ (Figure \ref{m_vs_ns}). We have discussed that larger gravity waves are produced via increasing $\alpha$. This effect is evident from Figures \ref{m_vs_ns} and \ref{ns_vs_kappa}, which show that $m$ and $\kappa$ values are raised as $\alpha$ increases, particularly in red-tilted regions. Figure \ref{runFig} depicts the absolute value of the running of $n_s$, $\log_{10} \vert \frac{\text{d} n_s}{\text{d} \ln k} \vert$. In our model, $\frac{\text{d} n_s}{\text{d} \ln k}$ is negative, and $\vert \frac{\text{d} n_s}{\text{d} \ln k} \vert \sim \mathcal{O}(10^{-3})-\mathcal{O}(10^{-4}) $.

\section{Conclusion}\label{conclusion}
In this letter we have considered the effects of the $R$-symmetry violating $\frac{\alpha}{m_P} {S^4}$ term in the standard hybrid inflationary scenario. By allowing for $R$-symmetry violation beyond the renormalizable level, we can give masses to right-handed neutrinos and down-type quarks in a simple way within flipped $SU(5)$, providing a particularly well-motivated inflationary scenario. We compare our results to the WMAP nine-year parameters $n_s$ and $r$, finding that we can easily generate red spectral tilt values within the observational bounds. This type of model can generate larger gravity waves than the standard ($\alpha = 0 $) case; raising $\alpha$ from $10^{-14}$ to $10^{-9}$ corresponds to an increase in $r$ from $10^{-12}$ to $10^{-8}$ at the WMAP nine-year central $n_s$ value \cite{WMAPnine}. This result has been achieved using a minimal K\"ahler potential, positive TeV-scale soft mass squared terms, a negative linear and a negative quartic soft term, and SUGRA correction terms up to sixth order in $x$. Additional interesting results were observed in the parameter study, which yielded a region of qualitatively new ``horizontal solutions" in the red-tilted region, above the main $r-n_s$ curve in Figure \ref{r_vs_nsFIXED}. The parameter study yields $r$ values as large as $\sim 10^{-4}$ in the red-tilted region. Finally, we observe that $\vert \alpha \vert$ develops a numerical upper bound of $\sim 10^{-7}$, while $\vert \frac{\text{d} n_s}{\text{d} \ln k} \vert \lesssim \mathcal{O}(10^{-3})-\mathcal{O}(10^{-4})$. 

\begin{acknowledgments}
This work is supported in part by the Delaware Space Grant College and Fellowship Program (NASA Grant NNX10AN63H) (M.C.), the University of Delaware Summer High Performance Computing Fellowship sponsored by NSF grant OCI 0904934 (E.S.), DOE grant DE-FG02-12ER41808, and by the University of Delaware (J.W.). We wish to give special thanks to Bumseok Kyae for his insightful comments and suggestions. 
\end{acknowledgments}

\appendix

\renewcommand{\thesection}{\arabic{section}}
\renewcommand{\thesubsection}{\arabic{subsection}}

\section{Minimization of the Potential With Respect to $\theta_S$}\label{app1}
The phase-dependent terms in our potential are, including just the global SUSY and soft SUSY-breaking terms:
\begin{equation*}
	V(\theta) \equiv - c_1 \cos(\delta) + c_2 \left( 2 \cos(\theta_s) - \vert A \vert \cos(\psi) \right) + c_3 \left( \vert A \vert \cos(\delta + \psi) + \cos(\theta_s + \delta) \right),
\end{equation*}
where the phases and coefficients (the latter are dimensionful) are
\begin{gather*}
	\delta \equiv \theta_\alpha + 3 \theta_s \quad , \quad \psi \equiv \theta_s + \theta_A\\
	c_1 \equiv \frac{8 \vert \alpha \vert \vert S \vert^3 M \kappa}{m_P} \quad , \quad c_2 \equiv 2 m_{3/2} M^2 \kappa \vert s \vert \quad , \quad c_3 \equiv \frac{2 m_{3/2} \vert \alpha \vert \vert S \vert^4}{m_P}.
\end{gather*}

We seek to minimize this function with respect to to $\theta_s$, thus we impose the conditions
\begin{align*}
	\frac{\text{d} V_\theta}{\text{d} \theta_s} &= 3 c_1 \sin(\delta) + c_2 \left[ -2 \sin(\theta_s) + \vert A \vert \sin(\psi) \right] - 4 c_3 \left[\vert A \vert \sin(\delta + \psi) + 	\sin(\theta_s + \delta) \right] = 0, \\
	\frac{\text{d}^2 V_\theta}{\text{d} {\theta_s}^2} &= 9 c_1 \cos(\delta) + c_2 \left[ -2 \cos(\theta_s)+ \vert A \vert \cos(\psi) \right] - 16 c_3 \left[ \vert A \vert \cos(\delta + \psi) + \cos(\theta_s + \delta) \right] > 0 .
\end{align*}

The first condition is met trivially by setting
\begin{equation*}
	\sin(\theta_s) = 0 \quad , \quad \sin(\psi) = 0 \quad , \quad \sin(\delta + \psi) = 0 \quad , \quad \sin(\theta_s + \delta) = 0,
\end{equation*}
which implies that each argument is an integer multiple of $\pi$. Finally, the second condition can be satisfied with
\begin{equation*}
	\cos(\theta_s) < 0 \quad , \quad \cos(\psi) > 0 \quad , \quad \cos(\delta) > 0 \quad , \quad \cos(\theta_s + \delta) < 0 \quad , \quad \cos(\delta + \psi) > 0.
\end{equation*}

The last inequality is a consequence of the previous four, but we can nonetheless ensure that the second condition is satisfied by declaring $\vert A \vert < 1$, which, from Equation \eqref{softcoefficients}, is equivalent to $b < 0$. With these phases the linear coefficient is $a = -2(2 +\vert A \vert) < 0$. Finally, the only phase-dependent SUGRA correction term we are including is $V_{\theta;\text{SUGRA}} = -\frac{12 \kappa M^2 \vert M \vert^5 \vert \alpha \vert }{{m_P}^3} \cos(\delta)$, which will always be negative with our chosen phases. 

 
\section{Blue-Tilt in Standard Global Plus $S^4$ Case}\label{app2}
It is interesting to investigate whether inflation can be driven exclusively using the new $S^4$ contribution without extending to SUGRA.  In global SUSY, the scalar potential acquires two new terms (see Equation \eqref{VGlobal}) which impart a nonzero slope along the inflationary valley.  For $|\alpha|$ small enough to constitute a perturbation, the $|s|^6$ term is strongly subdominant to the $|s|^3$ term, and thus the latter is chiefly responsible for meaningful alterations to the inflationary dynamics.

Some care must be taken in choosing the phases of the complex quantities in this version of the model.  The new term in the superpotential induces an additional SUSY vacuum appearing along the $|\phi| = 0$ direction. This vacuum is gauge symmetric and must be avoided if the symmetry breaking structure of the model is to be preserved.  If the phase factor $\cos( \theta_\alpha + 3 \theta_S)$ is positive, a stable inflationary valley along $|\phi| = 0$ leads to this vacuum.  Therefore we require this phase factor to be negative.  To achieve this, one (but not both) of $\theta_\alpha, \theta_S$ must be an odd integer multiple of $\pi$.  For simplicity, we choose $\theta_S = 0, \theta_\alpha = n\pi$ ($n$ odd).  Additionally, these phases result in a slight ($\sim$cubic) uplifting of the potential, consistent with a perturbation to the original hybrid model.

With these phases chosen, we examine the constraint imposed on the parameters of the model by the slow roll conditions $\epsilon_0 < 1, \eta_0 < 1$.  We find that the constraint from $\eta_0$ is convincingly more stringent than that from $\epsilon_0$, and so we will concentrate on the former.  In terms of dimensionful quantities, we may write analytically
\begin{eqnarray*}
	\eta_0 &=& 24 m_P |\alpha| |s_0| \left[ \frac{\kappa M^2 + 10 |\alpha| |s_0|^3 / m_P}{\left(\kappa M^2 + 4 |\alpha| |s_0|^3 / m_P\right)^2} \right] , \\
	&\simeq&  \frac{24 m_P |\alpha| |s_0|}{\kappa M^2 + 4 |\alpha| |s_0|^3 / m_P} .
\end{eqnarray*}
The denominator of this expression is strikingly close to Equation \eqref{critical} specifying the waterfall point $\tilde{s_c}$, after adjusting for phase differences (which amounts to a sign flip for the $\alpha$-dependent term).  Approximating $|s_0| \approx \tilde{s_c}$, we have
\begin{equation}
	\eta_0 \simeq \frac{24 m_P |\alpha| |s_0|}{\kappa \tilde{s_c}^2} .
\end{equation}
Taking $\eta_0 < 1$ and rearranging, we may cast the slow roll condition in the form
\begin{equation}
	\frac{|\alpha|}{\kappa} \lesssim \frac{1}{24} \left( \frac{|s_0|}{m_P} \right) \left( \frac{\tilde{s_c}}{|s_0|} \right)^2 .
\label{eta_analytical}
\end{equation}

Next, we investigate the condition for obtaining a red-tilted spectrum in this model.  Since $\eta_0 > 0$ in the present case, $n_s < 1$ requires $6 \epsilon_0 > 2 \eta_0$ according to Equation \eqref{r_ns}, or
\begin{equation*}
	\frac{\epsilon_0}{\eta_0} > \frac{1}{3} .
\end{equation*}
Using the same approximation stated above, we may write
\begin{equation*}
	\epsilon_0 \simeq 144 \left( \frac{\alpha}{\kappa} \right)^2 \left( \frac{|s_0|}{\tilde{s_c}} \right)^4 .
\end{equation*}
Combining this with Equation \eqref{eta_analytical} yields a red tilt condition
\begin{eqnarray}
	\frac{|\alpha|}{\kappa} \left( \frac{|s_0|}{\tilde{s_c}} \right)^2 \left( \frac{|s_0|}{m_P} \right) &>& \frac{1}{18} \sim 10^{-1} , \nonumber \\
	\frac{|\alpha|}{\kappa} &\gtrsim& 10^{-1} \left( \frac{m_P}{\tilde{s_c}} \right) \left( \frac{\tilde{s_c}}{|s_0|} \right)^3 .
\end{eqnarray}

Taking $1/24 \sim 10^{-1}$ in Equation \eqref{eta_analytical}, the slow roll and red tilt conditions may be consolidated into one expression,
\begin{equation}\label{SR+RT}
	10^{-1} \left( \frac{m_P}{\tilde{s_c}} \right) \left( \frac{\tilde{s_c}}{|s_0|} \right)^3 \lesssim \frac{|\alpha|}{\kappa} \lesssim 10^{-1} \left( \frac{|s_0|}{m_P} \right) \left( \frac{\tilde{s_c}}{|s_0|} \right)^2 .
\end{equation}
The middle portion $|\alpha| / \kappa$ may safely be omitted, after which several common factors may be canceled from the inequality.  Simplifying Equation \eqref{SR+RT} readily leads to
\begin{equation}
	|s_0| \gtrsim m_P .
\end{equation}
In other words, the global-SUSY version of the model under the slow roll approximation may only lead to $n_s < 1$ for trans-Planckian values of the inflaton field.  Since the SUGRA sector is suppressed by powers of $|s_0| / m_P$, it is then inconsistent to treat the model only using global SUSY.

In addition to these arguments, the assumptions placed on the phases for this version of the model are also problematic.  Above, we commented that the phase factor $\cos( \theta_\alpha + 3 \theta_S)$ must be negative in order to preserve gauge symmetry breaking.  Referring to Appendix \ref{app1}, however, the potential is minimized with respect to $\theta_S$ for $\cos( \theta_\alpha + 3 \theta_S) > 0$.  Therefore an unattractive choice must be made: allow the phase $\theta_S$ to vary dynamically (potentially ruining inflation in the process), or sacrifice gauge symmetry breaking (a defining characteristic of hybrid inflation).  This issue compounds the difficulties in obtaining a red-tilted spectrum via slow roll.  Thus we conclude that successful inflation may not be implemented in our $S^4$ model using only global SUSY.

\bibliography{CRSSW_alpha}

\end{document}